\documentclass[prl,twocolumn]{revtex4}

\usepackage[dvips]{graphicx} \usepackage{amsmath}

\DeclareMathOperator{\erf}{erf} 
\newcommand{\avg}[1]{\ensuremath{\langle #1 \rangle}}
\newcommand{\D}{\ensuremath{\mathrm{d}}}
\newcommand{\rme}{\ensuremath{\mathrm{e}}}

\begin{document}

\title{Survival probability of a diffusing particle constrained by two 
moving, absorbing boundaries}
\date{\today}
\author{Alan J. Bray}
\author{Richard Smith}
\affiliation{School of Physics and Astronomy, The University of Manchester, 
Manchester M13 9PL, U.K.}

\begin{abstract}
We calculate the exact asymptotic survival probability, $Q$, of a 
one-dimensional Brownian particle, initially located at the point 
$x \in (-L,L)$, in the presence of two moving absorbing boundaries located 
at $\pm (L+ct)$. The result is $Q(y,\lambda) = \sum_{n=-\infty}^\infty 
(-1)^n \cosh(ny)\,\exp(-n^2\lambda)$, where $y=cx/D$, $\lambda = cL/D$ and 
$D$ is the diffusion constant of the particle. The results may be extended 
to the case where the absorbing boundaries have different speeds. As an 
application, we compute the asymptotic survival probability for the 
trapping reaction $A + B \to B$, for evanescent traps with a long decay 
time. 

\end{abstract}

\maketitle

Physical  systems  described by  partial  differential equations  with
moving  boundary  conditions are  ubiquitous  in nature  \cite{Crank}.
Unfortunately, such  equations are notoriously difficult  to solve.  The
case of  a single moving boundary  is often amenable  to analysis, for
example by tranforming to the moving  frame, but the case of more than
one moving boundary is, in general, intractable.

First-passage problems  are another field of research  for which there
are relatively few exact results \cite{Redner,Majumdar}.  The simplest
such  problem,  for which  some exact results are  available, is  that 
of a Brownian particle (i.e.\ random walker) moving in the  presence of 
one or  more absorbing  boundaries \cite{Redner}. The case of a single
boundary,   moving  at   constant   speed,  can   be  solved   exactly
\cite{Redner}  but, to our  knowledge, the  survival probability  of a
single Brownian walker  in the presence of two  moving boundaries with
different velocities had not been solved up to now.

In the  present paper we  apply ``backward Fokker Planck''  methods to
solve  this  problem exactly.   As  an  application,  we consider  the
one-dimensional trapping reaction $A + B \to B$, where the density of 
traps, $\rho$, decays exponentially  with time, and obtain the exact 
asymptotic form of the final $A$-particle density in the limit where the 
decay-time, $\tau$, of the traps is large.

\begin{figure}[h]
\includegraphics[width=0.8\linewidth]{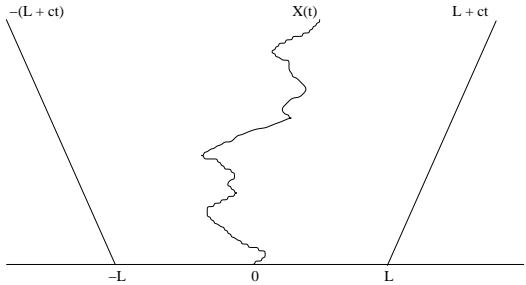}
\caption{Brownian  walker, starting from the origin, in  a  
linearly--expanding cage.}\label{expcgplot}
\end{figure}

We consider a Brownian walker moving according to the Langevin
equation $\dot{X}(t)=\eta(t)$, with initial condition $X(0)=x$, where 
$\eta(t)$  is Gaussian white  noise  with mean zero  and  correlator  
$\avg{\eta(t_1) \eta(t_2)}=2   D  \delta(t_1-t_2)$.  The  particle is  
flanked  by  two receding, absorbing  walls with coordinates  
$\pm (L+ct)$, as shown  in   Fig.~\ref{expcgplot}. We can derive  
a  backward Fokker-Planck  equation  for the  probability, $Q(x,L,t)$, 
that  the particle has  survived up to time $t$ having 
started  at position $x \in (-L,L)$. We note that, after 
infinitesimal time $\Delta t$, the particle will have have moved to 
position $x + \Delta X$, and the walls will have moved to positions 
$\pm(L + c\Delta t)$. It follows that 
$Q(x,L,t) = \langle Q(x+\Delta X, L + c \Delta t, t-\Delta t) \rangle$, 
where the average is over the distribution of the spatial increment 
$\Delta X$. Expanding to first order in $\Delta t$, using 
$\langle \Delta X \rangle = 0$ and $\langle (\Delta X)^2 \rangle 
= 2D \Delta t$, yields the backward Fokker-Planck equation
\begin{equation}
\label{bfpe}
\frac{\partial     Q}{\partial     t}=D\frac{\partial^2    Q}{\partial
x^2}+c\frac{\partial Q}{\partial L}.
\end{equation}

The infinite-time result is obtained by setting the time derivative to
zero. It is convenient to introduce the variables  
$y = cx/D$ and $\lambda = cL/D$ to represent the dimensionless initial 
positions of the particle and the walls. Eq.~(\ref{bfpe}) then reads, 
at infinite time,  
\begin{equation}
\label{rescale}
\frac{\partial^2     Q}{\partial    y^2}+\frac{\partial    Q}{\partial
\lambda}=0,
\end{equation}
where $-\lambda \le y \le \lambda$, subject  to  the absorbing 
boundary conditions, $Q(\pm\lambda,\lambda)=0$, and $Q(y,\infty)=1$, 
since if the particle starts at one of walls it is immediately absorbed,  
while if the walls are initially infinitely  far away the particle  
will survive with  probability $1$. A  solution of Eq.~(\ref{rescale}) 
satisfying the  boundary  conditions may  be  deduced  by  inspection, 
noting the symmetry of the problem under reflection, $y \to -y$:
\begin{equation}
\label{coshsoln}
Q(y,\lambda)=  \sum_{n=-\infty}^\infty (-1)^n\cosh (ny)\, \rme^{-n^2
\lambda}.
\end{equation}
Despite the simplicity of its derivation, Eq.\ (\ref{coshsoln}) is, to our 
knowledge, a new result. It should be noted that this result seems 
difficult to obtain using conventional (``forward'') Fokker-Planck methods.
The backward Fokker-Planck method has eliminated the time-dependent 
boundary conditions from the problem at the cost of introducing 
the initial wall parameter, $L$, as an additional independent variable.  

\begin{figure}[h]
\includegraphics[width=\linewidth]{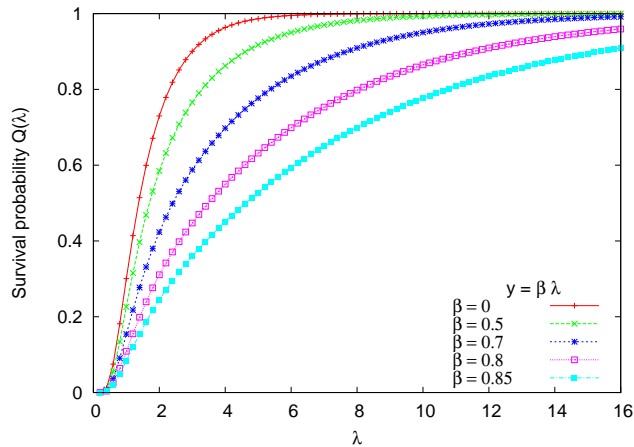}
\caption{Survival probability $Q(\lambda)$ for various values of the    
initial coordinate $y$.}
\label{xcgyplot}
\end{figure}

In  Fig.~\ref{xcgyplot} we present the results of numerically evaluating 
the sum in Eq.\ (\ref{coshsoln}) for various values of the starting 
coordinate, $y$. 

For  a particle  starting at  the origin, the survival probability 
$Q(0,\lambda)$, written simply as $Q(\lambda)$,  is
\begin{equation}
\label{qlambda}
Q(\lambda)= \sum_{n=-\infty}^\infty (-1)^n \rme^{-n^2 \lambda}.
\end{equation}
To leading order for large $\lambda$ we may approximate the sum by the
$n=0$ and $n=\pm 1$ terms:
\begin{equation}
\label{largelambda}
Q(\lambda) \sim 1-2\rme^{-\lambda }.
\end{equation}
This result agrees,  to first order in ${\rme}^{-\lambda}$, with the  
approximate infinite-time  result, $Q(\lambda) \sim \exp(-2\rme^{-\lambda})$, 
obtained using the method of Krapivsky and Redner \cite{krap} for a  
rapidly expanding  cage. 
In this approach one writes the joint probabability distribution that 
the particle survives till time $t$, and is located at $x$, 
in the approximate form \cite{krap} $P(x,t) = [Q(t)/\sqrt{4\pi Dt}] 
\exp(-x^2/4Dt)$, i.e.\ one multiplies the free diffusion propagator 
by the probability, $Q(t)$, for the particle to survive till time $t$, 
ignoring the boundary conditions at the walls, where the density is 
anyway small. The rate of change of $Q(t)$ is minus the total 
probability flux through the walls, 
$dQ/dt = 2 D(\partial P/\partial x)_{x=L+ct}$. Integrating the resulting 
equation from $t=0$ to $t=\infty$, using the method of steepest decents  
(valid for $\lambda \gg 1$), gives the quoted result. We see that this 
method only gives the leading departure from unity correctly.

For small $\lambda$, the leading-order behaviour can be obtained by 
rewriting Eq.~(\ref{qlambda}) using the Poisson sum formula,
$\sum_{n=-\infty}^\infty   f(n)=\sum_{k=-\infty}^\infty  \tilde{f}(2\pi
k)$, where  $\tilde{f}(k)$ is  the Fourier  transform of  the  the function
$f(n)$. This gives
\begin{equation}
\label{poisson}
Q(\lambda)=\sqrt{\frac{\pi}{\lambda}}\sum_{k=-\infty}^\infty
\rme^{-\pi^2 (2k-1)^2 /4 \lambda}
\end{equation}
The leading behavior at small $\lambda$ is 
\begin{equation}
Q(\lambda) \sim 2\sqrt{\pi/\lambda}\,\exp(-\pi^2/4\lambda),
\label{poisson_leading} 
\end{equation}
which contains an essential singularity at $\lambda=0$. We are unaware
of any approximate methods for which even the leading  behavior, 
Eq.~(\ref{poisson_leading}), can be recovered. 

\begin{figure}[h]
\includegraphics[width=\linewidth]{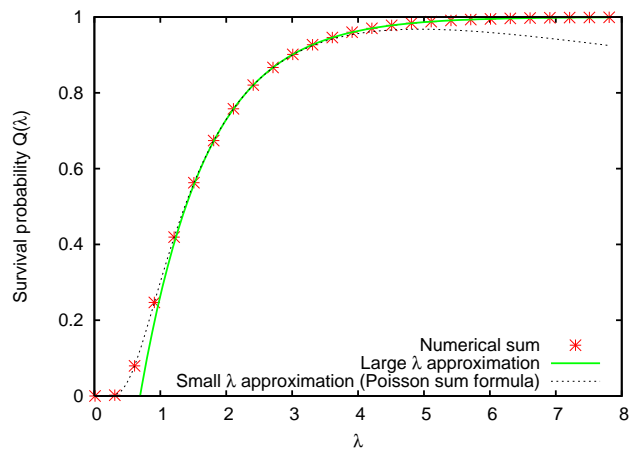}
\caption{Survival  probability for a particle starting at the mid-point of  
a linearly-expanding  cage with the asymptotic forms for small and large 
values of $\lambda$, the dimensionless measure of the initial locations
of the walls.} 
\label{xcglambdaplot}
\end{figure}

We have performed the summ in Eq.~(\ref{qlambda}) numerically and plot 
the result together with the  large and  small $\lambda$ forms, 
Eq.~(\ref{largelambda}) and Eq.~(\ref{poisson_leading}), in  
Fig.~\ref{xcglambdaplot}.
We see that the asymptotic forms describe the data well over a considerable 
range of $\lambda$. This is readily understood on noting that the leading 
corrections to the asymptotic forms are of order $\exp(-4\lambda)$ and 
$\exp(-9\pi^2/4\lambda)$ for large and small $\lambda$ respectively.

To discuss the general case of walls with different speeds, we first 
consider the problem of a Brownian walker with drift $\alpha c$
following a path  $X(t)$, $X(0)=x$, with an absorbing  boundary at the
origin,  and a receding  absorbing boundary at position $L+ct$,  where 
$x \in (0,L)$. The  path $X(t)$ of  the walker satifies the Langevin
equation  $\dot{X}(t) = \eta(t)+\alpha  c$,  where  $\eta(t)$  is  again
Gaussian  white noise.  We  can write  down a  Backward Fokker-Planck
equation  for the  survival probability  of the  particle  at infinite
time which, making the  same changes  of variable  as those  used in
Eq.~(\ref{rescale}), takes the form
\begin{equation}
\label{driftfokker}
\frac{\partial^2  Q}{\partial  y^2}+\alpha \frac{\partial  Q}{\partial
y}+ \frac{\partial Q}{\partial \lambda}=0,
\end{equation}
subject     to      the     boundary     conditions     $Q(y=0,\lambda
)=Q(y=\lambda,\lambda)=0$. In the limit $\lambda \to \infty$ we impose
as a boundary condition  the solution of Eq.~(\ref{driftfokker}) with
$\partial Q/\partial \lambda=0$, i.e.\ $Q(y,\lambda \to \infty)=1-
\exp (-\alpha  y)$, which is the survival probability of a diffusing 
particle that drifts with mean velocity $\alpha c$ away from a fixed 
absorbing boundary or, equivalently, the survival  probability of  a particle
with no drift  in the presence of one wall  receding at speed $\alpha c$
(see, for example, ref.~\cite{Redner} and references therein).

A  solution to  Eq.~(\ref{driftfokker}) satisfying  all  the boundary
conditions is
\begin{equation}
\label{driftsoln}
Q(y,\lambda)=\sum_{n=-\infty}^\infty \left( \rme^{ny}-\rme^{-(n+\alpha
)y} \right) \rme^{-n(n+\alpha) \lambda}.
\end{equation}
On  shifting  to the  left  both  the path  of  the  particle and  the
coordinates of  the barriers by  a distance $(L/2 +\alpha ct)$,  we see
that  Eq.~(\ref{driftsoln}) solves  the expanding  cage  problem for 
a particle with no drift, and with
asymmetrically  receding  walls---the  left  wall  having  coordinates
$-(L/2 +\alpha ct)$,  and the right  $(L/2 +(1-\alpha )ct)$, and  may be
shown  to  be equivalent  to  Eq.~(\ref{coshsoln})  when $\alpha = 1/2$
(noting that the separation of  the walls and drift velocities need to
be doubled to show an exact mapping between the two problems).

As an  application of  these results, we consider the trapping 
reaction, $A + B \to B$ \cite{ovchinnikov}, with evanescent traps. 
We find  the infinite-time
survival probability of a particle, $A$, surrounded by an infinite sea
of  Poisson-distributed, evanescent traps,  $B$, with  initial density 
$\rho$, undergoing   the trapping reaction  in  one  dimension.  By
Poisson-distributed we  mean that, at $t=0$, each  infinitesimal interval 
$dx$ contains  a  trap  with  probability  $\rho dx$.  In  particular,  the
probability to  find no traps  in an interval  of length $L$  is $\exp
(-\rho L)$.  The traps are  evanescent in the sense that they randomly
and independently disappear from the system in such a way that the
overall  trap density  decreases  in a  prescribed fashion,  $\rho(t)=
f(t)\rho(0)$. The particle and traps both perform Brownian motion, 
with diffusion coefficients $D_A$ and $D_B$ respectively. This problem
has recently  been studied in  detail for various functions $f(t)$ for 
the  case of a  fixed target ($D_A=0$) \cite{lindenberg}. It was shown 
that, in this case, there is a non-zero infinite-time survival probability 
whenever $f(t)$ falls off more rapidly that $t^{-1/2}$ for large-$t$.  
Here we address the general and more difficult problem of a moving 
$A$-particle, and specialize to the case of exponentially decaying trap 
density, $f(t) = \exp(-t/\tau)$, anticipating a non-zero infinite-time 
survival probability. This survival probability for a single $A$-particle 
also gives the {\em fraction} of $A$ particles that survive if the initial 
state contains a macroscopic number of them.   

We approach the problem in the spirit of  Bray  and Blythe~\cite{bounds},  
who  considered  the trapping  reaction without trap  decay,  by finding  
upper and  lower bounds on the  survival   probability   and   showing  
that   they asymptotically agree.  In the calculation  of both bounds we  
use the formalism introduced  by Bray, Majumdar  and Blythe~\cite{canon}. 
They define a  quantity $\mu(t)$, the mean  number of different traps 
that would meet the $A$-particle up to time $t$ , for a given $A$-particle 
trajectory  $z(t)$, in a fictitious model where the $A$ and $B$ particles 
do not react. It satisfies the integral equation
\begin{equation}
\label{fundamental}
\rho(t) =\int_0^t \D t'\;\dot{\mu}(t')\;G\left( z(t),t | z(t'), t'\right),
\end{equation}
where $\rho(t)$ is the trap density, and $G\left( z(t),t | z(t'),t'\right)
= \exp \left[-(z(t)-z(t'))^2/4   D_B  (t-t') \right]/
\sqrt{4\pi D_B (t-t')}$ is the trap diffusion propagator.  
Note  that  $\mu [z]$  is  implicitly  a  functional  of  the
$A$-particle   trajectory,  and   that   $\mu(t=0)=0$.  The   survival
probability of the $A$ particle is then given by \cite{canon,anton}
\begin{equation}
\label{expmu}
Q(t)=\avg{\rme^{-\mu[z]}}_z,
\end{equation}
with the average taken over the $A$-particle trajectories with the usual 
Wiener measure. When the fraction of surviving traps is $f(t)$,
Eq.~(\ref{fundamental}) is modified thus \cite{canon}:
\begin{equation}
\label{fundamentaldecay}
\rho =\int_0^t \D t'  \; \frac{\dot{\mu}(t')}{f(t')} \; G\left(
z(t),t | z(t'),t' \right),
\end{equation}
where we  now use $\rho$  to refer to  the {\em initial} trap  density. 
This minor  modification means  that we  may simply  define a  new quantity
$\dot{\phi}(t)=\dot{\mu}(t)/f(t)$   in  Eq.~(\ref{fundamentaldecay}),
solve as in the non-decaying case, and find that
\begin{equation}
\label{oldnew}
\mu (t)=\int_0^t \D t' \; \dot{\phi}(t') f(t').
\end{equation}

We  first derive  an exact upper  bound on  the  survival  probability,  
as  in \cite{bounds,canon}, by arguing that the particle will survive 
longest if it remains  stationary at  the  origin (the `target problem' 
\cite{blumen}).  This  is  the so-called  
`Pascal principle' \cite{moreau}. We solve Eq.~(\ref{fundamentaldecay}) 
with $z(t)=0$,
\begin{equation}
\label{target}
\rho  =\int_0^t \D t'\;\frac{\dot{\phi}(t')}{\sqrt{4\pi  D_B(t-t')}},
\end{equation}
and get the solution found in \cite{bounds}, that is $\phi = 4 (\rho^2
D_B t/\pi)^{1/2}$ which, using Eq.~(\ref{oldnew}) with $t=\infty$ and 
$f(t') = \exp(-t'/\tau)$, and Eq.\ (\ref{expmu}), gives an  upper bound 
on  the eventual  survival probability as
\begin{equation}
\label{upperbound}
Q \leq \exp[-2 (\rho^2 D_B \tau)^{1/2}].
\end{equation}
We can  prove that  this  is a  rigorous  upper  bound following  the
procedure outlined in \cite{canon}.  If we write $\phi =\phi_0+\phi_1$
in  Eq.~(\ref{fundamentaldecay}), where $\phi_0$  is the  solution to
Eq.~(\ref{target}), we can show that  $\phi_1 \geq 0$,  proving that
$\phi \geq \phi_0$.

In \cite{bounds}, Bray and  Blythe bound the survival probability from
below by considering a notional  box centred on the origin, from where
the  $A$-particle's  trajectory is  chosen to  begin. They then select 
a subset of surviving trajectories by imposing  three independent 
conditions that together guarantee the $A$-particle's survival:  
(i) The $A$ particle does not leave  the box up to time $t$; (ii) no trap 
enters  the box up  to time $t$; and (iii) no traps were inside the box  
at time $t=0$.  These conditions undercount the number  of possible 
surviving  trajectories of the $A$-particle -- for example,  a trap may 
enter  the box  without  trapping the particle -- so the  probability 
of  fulfilling them  underestimates the actual survival probability. 
We follow a similar line of argument here, but
with a modification to adjust for the exponentially decaying traps: We
allow the walls of  the box to recede linearly. Then  we may use the
result obtained in Eq.~(\ref{qlambda}) to determine the probability 
of satisfying  condition (i) above.  We  consider, therefore, the survival 
at  infinite time. The probability corresponding to condition (iii) follows  
simply from the Poisson property of the trap density,  and for a box of 
length $2L$ is simply $\exp(-2\rho L)$.

To  obtain the  probability that  no traps  have entered the  box, we
consider each side independently and  square the result.  This is then
equivalent to the  survival probability of a ballistic $A$  particle in an
infinite  sea of  traps,  which was  solved in~\cite{ballistic}.   The
problem    may    be    solved    using    a    modified    form    of
Eq.~(\ref{fundamentaldecay}) to  allow for traps on only  one side of
the  wall, a  modification  introduced in~\cite{canon}. With  the
linear trajectory $z(t)=ct$ one obtains
\begin{equation}
\label{oneside}
\frac{\rho}{2}  \left[ 1+\erf  \left( \frac{c  \sqrt{t}}{\sqrt{4D_B}} \right)
\right]  = \int_0^t \D t'\,\dot{\phi}(t')  \frac{\exp \left(-
\frac{c^2}{4D_B}(t-t')\right)}{\sqrt{4\pi D_B (t-t')}}.
\end{equation}
We  solve this  for $\dot{\phi}(t)$  by Laplace  transform  and, using
Eq.~(\ref{oldnew}), obtain the probability
\begin{equation}
\label{label}
\exp  \left( -2\rho \sqrt{D_B  \tau+\frac{c^2 \tau^2}{4}}  -\rho c\tau
\right),
\end{equation}
that no traps have ever entered the box.

We now combine  these three factors to obtain a lower bound on 
$Q$: 
\begin{equation}
\label{lowerbound}
Q\geq \exp \left( -2\rho  \sqrt{D_B\tau + \frac{c^2 \tau^2}{4}}-\rho c
\tau-2\rho L-\frac{\pi^2 D_A}{4cL} \right),
\end{equation}
where the last term comes from using the small-$\lambda$ result 
(\ref{poisson_leading}) and we   have   neglected a 
logarithmic correction in the exponent coming from the pre-exponential 
factor in Eq.\ (\ref{poisson}). The result  is valid for  
$cL/D_A \ll 1$ (corresponding to small  $\lambda$ in the rescaled 
coordinates of Eqs.~(\ref{rescale})--(\ref{poisson_leading})).  This  bound 
contains two free parameters,  $L$  and  $c$, so  we  obtain  the  
best lower  bound  by maximizing  Eq.~(\ref{lowerbound})  with  respect  
to both  of  these quantities, giving
\begin{equation}
\label{optimal}
Q\geq \exp \left(-2(\rho^2 D_B \tau)^{1/2}- \frac{3}{2}(4\pi^2 \rho^2
D_A \tau)^{1/3} \right).
\end{equation}
This is  valid as long  as $\rho^2 D_A  \tau \gg 1$, and  $\rho^2 D_A
\tau \gg  (D_A/D_B)^3$, which are satisfied for large enough $\tau$ for 
any values of $\rho$, $D_A$, and $D_B$.  

Comparing the two bounds, (\ref{upperbound}) and (\ref{optimal}), we see 
that they agree, for large $\tau$, to leading order in the exponent. They 
therefore determine the exact large $\tau$ asymptotics in the form 
$Q \sim \exp[-2(\rho^2 D_B \tau)^{1/2} + \cdots]$, where the ellipsis 
indicates subdominant terms. Note that the leading term is independent of 
$D_A$, just as was found for the time-dependent asymptotics of the survival 
probability for non-decaying traps \cite{bounds}.

The fact that the bounds pinch indicates that the choice of linearly receding 
walls to obtain the lower bound is optimal for the case of exponentially 
decaying traps with a long decay time. Other forms for the trap decay 
function $f(t)$ will require a different choice for the wall motion to 
optimize the bound. We recall that for non-decaying traps the bound is 
optimal for static walls \cite{bounds}. 

In summary, we have obtained exact results for the infinite-time survival 
probability of a diffusing particle flanked by two receding, absorbing 
boundaries, including the case where the boundaries move at different speeds.
We have used the results to compute the survival probability of a 
particle diffusing in a sea of diffusing, evanescent traps in the limit 
where the trap decay time is large. Extensions to systems with general 
spatial dimensionality are possible and will be discussed elsewhere. 

The work of RS was supported by EPSRC (UK).

\end{document}